# Room Temperature Thermoelectric Properties of Isostructural Selenides $Ta_2PdS_6$ and $Ta_2PdSe_6$

Akitoshi Nakano[1]*, Urara Maruoka[1], Fumiaki Kato[1], Hiroki Taniguchi[1] and Ichiro Terasaki[1]

[1]*Department of Physics, Nagoya University, Nagoya 464-8602, Japan*

We have measured thermoelectric properties of $Ta_2PdX_6$ (X=S, Se) around room temperature using single crystal samples. We find that the power factor of $Ta_2PdX_6$ is relatively high from middle-low to room temperatures, and notably $Ta_2PdSe_6$ shows the largest power factor among thermoelectric materials with an electrical conductivity of $10^{-2}$ Ωcm at 300 K. $Ta_2PdS_6$ will be a possible candidate for a Peltier cooling material if the lattice thermal conductivity is reduced by chemical substitution.

Transition-metal chalcogenides have attracted much attention due to their unique physical properties arising from their low-dimensionality. Charge density wave[1], superconductivity[2], and topologically nontrivial electronic phases[3] in binary transition-metal dichalcogenides have been one of the central issues of the condensed matter physics. Ternary transition-metal chalcogenides with strong one-dimensionality (1D) also attract keen interests due to their exotic physical properties. For instance, a quasi 1D telluride $Ta_4SiTe_4$ and its sister compounds show good thermoelectric properties from middle-low to room temperatures[4]-[6]. Another quasi 1D selenide $Ta_2NiSe_5$ is a candidate of the excitonic insulator (EI)[7]-[9], which has been pursued for long time from its theoretical prediction in the 1960s.

$Ta_2PdX_6$ (X=S, Se), which was synthesized in 1985[10], is yet another 1D ternary transition-metal chalcogenide. This compound crystallizes in the monoclinic structure with the space group $C2/m$. $Ta_2PdX_6$ has a layered crystal structure, where the square planner $PdX_4$ and face-shared $TaX_6$ prisms form 1D chains along the *b*-axis direction as shown in Fig.1(a). Except for superconductivity with $T_c$ = 4.5 K in $(C_2H_8N_2)_yTa_2PdSe_6$[11] and ultrahigh photo-responsibility in atomic layer $Ta_2PdS_6$[12], fundamental bulk transport properties of $Ta_2PdX_6$ have not been investigated yet.

In this letter, we focus on the transport properties of $Ta_2PdX_6$. The chemical formula and the characteristic 1D crystal structure are quite similar to the EI candidate $Ta_2NiSe_5$, which we have investigated by means of structural analyses[13,14] and transport measurement[15,16] thus far. This structural similarity suggests that $Ta_2PdX_6$ may be on the verge of the excitonic instability. Since the EI is predicted to realize in a semiconductor or semimetal, the bandgap $|E_g|$ of which are smaller than the binding energy between an electron and a hole[17-22], the transport measurements are essential to explore EIs. In this context, we conducted transport measurements on $Ta_2PdX_6$ to compare with $Ta_2NiSe_5$. Although no reminiscent signs of an excitonic phase transition was observed in $Ta_2PdX_6$, we instead found that $Ta_2PdX_6$ shows relatively large thermoelectric power factor from middle-low to room temperatures.

High-quality single crystals of $Ta_2PdX_6$ (X = S, Se) were grown by means of $I_2$ vapor transport. Powders of tantalum (99.9%), palladium (99.9%), and sulfur/selenium (99.999%) were loaded into an evacuated quartz tube with an $I_2$ concentration of ~3 mg/cm$^3$. Then, a temperature difference of 145 °C between 875 °C and 730 °C in a three-zone furnace was used for crystal growth for 4 days. Shiny silver needle-like single crystals of $Ta_2PdX_6$ (Figs.1(b) and (c)) are obtained at the cold end of the tube. The crystals are identified by single crystal X-ray diffraction. We confirmed that the needle direction is along the crystal *b*-axis.

Transport properties along the *b*-axis direction, including resistivity, thermopower, and Hall resistivity, were measured using PPMS (Quantum Design). The resistivity was measured by a four-probe method using gold wires with 20 μm diameters and silver paste. The thermopower was measured with a steady state and the two-probe technique. The sample bridged two separated copper heat baths, and the resistance heater created a temperature difference between the two heat baths, which was monitored through a copper-constantan differential thermocouple. The Hall resistivity was measured by sweeping magnetic field from −7 to 7 T at constant temperatures. The signal was collected using ΔR mode of a nano-ohmmeter LR-700 (LINEAR RESEARCH INC). In-plane thermal diffusivity was measured with a home-made measurement station based on an ac calorimetric method. The details will be published in a separate paper.

Figure 2(a) shows the temperature dependence of the resistivity and the Seebeck coefficient of $Ta_2PdS_6$. The resistivity is relatively low, 8 mΩcm at 300 K. As temperature decreases, the resistivity exponentially increases. The activation energy estimated from the Arrhenius plot is ~40 meV, indicating a narrow gap semiconductor state. The Seebeck coefficient is negative and as large as 450 µV/K at 300 K. The absolute value of Seebeck coefficient increases to 600 µV/K as temperature decreases down to 100 K.

The inset of Fig.2(a) shows the field dependence of the Hall resistivity at 100 K. The value of the Hall coefficient at 100 K is -1.4 $cm^3$/C. The negative sign is consistent with the sign of the Seebeck coefficient. If we assume a single carrier model, the estimated electron density and carrier mobility $R_H/\rho$ at 100 K are a low value of $4.6 \times 10^{18}$ $cm^{-3}$ and a relatively high value of 60 $cm^2$/Vs, respectively.

Figure 2(b) shows the temperature dependence of the resistivity and the Seebeck coefficient of $Ta_2PdSe_6$. In contrast to the semiconducting resistivity of $Ta_2PdS_6$, $Ta_2PdSe_6$ shows metallic resistivity which decreases linearly upon cooling. The drastic change from semiconductor to metal depending on iso-valent X may imply that the band structure near the Fermi level on $Ta_2PdX_6$ is significantly affected by the difference of electronegativity between sulfur and selenium. At room temperature, the sign of the Seebeck coefficient is negative and smaller than the Seebeck coefficient for $Ta_2PdS_6$. As temperature decreases, the Seebeck coefficient goes to zero around 100 K, indicating the sign change.

The field dependence of the Hall resistivity at 120 K is shown in the inset of Fig. 2(b). Contrary to $Ta_2PdS_6$, the sign of the Hall coefficient at 120 K is positive, indicating coexistence of electrons and holes. If we employ a two carrier model, the Hall coefficient $R_H$ can be written as

$$R_H = \frac{p\mu_h^2 - n\mu_e^2}{e(p\mu_h + n\mu_e)^2}. \qquad (1)$$

where, $n$ ($p$), $e$, and $\mu_{e(h)}$ are the concentration of electrons (holes), the element charge, and the carrier mobility of the electron (hole), respectively. Imposing a semimetallic condition $n = p$, we can rewrite equation (1) as

$$R_H = \frac{\mu_h - \mu_e}{ne(\mu_h + \mu_e)}. \qquad (2)$$

Since the resistivity can be written as

$$\rho = \frac{1}{ne(\mu_h + \mu_e)}, \qquad (3)$$

using equation (2) and (3) we can obtain the difference between the carrier mobility of electrons and holes as

$$R_H/\rho = \mu_h - \mu_e \qquad (4)$$

Using equation (4) $R_H/\rho$ is estimated to be ~ 400 cm$^2$/Vs, indicating a high and asymmetric carrier mobility in $Ta_2PdSe_6$.

Figure 2(c) shows the thermoelectric power factor of $Ta_2PdX_6$ as a function of temperature. The power factor $S^2/\rho$ is a measure of the electric power converted from thermal energy across a temperature difference of 1 K. Thus, a material with higher power factor is demanded to generate more electricity. We also plot the power factor of an EI candidate $Ta_2NiSe_5$[15], together with pristine $Bi_2Te_3$[23] that is a commercially available thermoelectric material. As shown in Fig. 2(c), $Ta_2PdX_6$ shows higher power factor than $Ta_2NiSe_5$ in spite of the structural similarity. Furthermore, the power factor of $Ta_2PdS_6$ is larger than pristine $Bi_2Te_3$ in a wide temperature range from 100 to 300 K.

Figure 3 shows the room temperature power factor of various materials[24] plotted as a function of resistivity. We find that the power factor of $Ta_2PdX_6$ is in a top level among thermoelectric materials reported thus far, and especially $Ta_2PdS_6$ shows the largest power factor among the materials which have resistivity around $10^{-2}$ Ωcm. In the case of semimetallic $Ta_2PdSe_6$, the relatively large power factor stems from the high electric conductivity due to high carrier mobility, whereas in the case of semiconductor $Ta_2PdS_6$, the large power factor comes from the large Seebeck coefficient due to existence of the bandgap.

Note that the power factor of $Ta_2NiSe_5$ is three orders magnitude smaller than that of $Ta_2PdS_6$ at 300 K. Since their resistivity is of the same order, the small power factor of $Ta_2NiSe_5$ is due to the smaller Seebeck coefficient of $Ta_2NiSe_5$ than $Ta_2PdS_6$. The smaller Seebeck coefficient in spite of the larger bandgap ~160 meV[9] indicates that the significant compensation between the Seebeck coefficient of electrons and holes. According to the recent Hall resistivity measurements, the difference between the carrier mobility of electrons and holes of $Ta_2NiSe_5$ is smaller than 20 cm$^2$/Vs[25] in a wide temperature range, whereas $Ta_2PdS_6$ shows a larger $R_H/\rho$ of 60 cm$^2$/Vs. This suggests that the small difference between the carrier mobility of electrons and holes plays a role for the compensation of the Seebeck coefficient in $Ta_2NiSe_5$. Instead, the almost equal carrier mobility of electrons and holes may favor the excitonic bound state of electrons and holes in $Ta_2NiSe_5$.

Finally, we evaluate the dimensionless figure of merit $ZT$ ($= S^2/\rho\kappa$) of $Ta_2PdX_6$ which is a measure of energy-conversion efficiency. Combining the thermal diffusivity measured by ac calorimetric method with the heat capacity assumed by the Dulong-Petit law, $\kappa$ is determined to be 14 W/mK for $Ta_2PdS_6$ and 17 W/mK for $Ta_2PdSe_6$, as summarized in Table.1. Employing these $\kappa$ values, we evaluate $ZT$ to be $6.4\times10^{-2}$ and $2.1\times10^{-2}$, for $Ta_2PdS_6$ and $Ta_2PdSe_6$, respectively, which are as large as other sulfides[26] and far below $Bi_2Te_3$ for $ZT = 1$[27]. This is due to 15 times higher $\kappa$ of $Ta_2PdX_6$ than that of $Bi_2Te_3$[27]. Applying the Wiedemann-Franz law to the $\rho$ data, the electron contribution to the thermal conductivity is negligible for $Ta_2PdS_6$. The evaluated phonon mean free path at 300 K for $Ta_2PdS_6$ is about 5 ~ 10 nm which is longer than the lattice constant. Thus, if we can simultaneously suppress the lattice thermal conductivity and resistivity by chemical substitution, $Ta_2PdS_6$ can be a potential thermoelectric material.

We conducted transport measurements on $Ta_2PdX_6$ (X=S, Se) which has a similar characteristic one-dimensional crystal structure to an excitonic insulator candidate $Ta_2NiSe_5$. From the resistivity and the Seebeck coefficient measurements, we reveal that $Ta_2PdS_6$ is a semiconductor with a narrow gap of 40 meV, whereas $Ta_2PdSe_6$ is a good metal. The thermoelectric power factor is relatively large for the both compounds. Especially, $Ta_2PdS_6$ shows the largest power factor among the materials which have resistivity around $10^{-2}$ $\Omega$cm. The evaluated dimensionless figure of merit of $Ta_2PdX_6$ is as large as other sulfides. If we can simultaneously suppress the lattice thermal conductivity and resistivity by chemical substitution, $Ta_2PdS_6$ can be a potential thermoelectric material.

**Acknowledgments**

This work was partly supported by Nanotechnology Platform Program (Molecule and Material Synthesis) of the Ministry of Education, Culture, Sports, Science and Technology (MEXT), Japan, Grant Number JPMXP09S20NU0029. This work is partially supported by Kakenhi Grant Nos. 17H06136, 19H05791 and 20K20898 of Japan.

*E-mail: nakano.akitoshi@f.mbox.nagoya-u.ac.jp

Fig.1

(a)

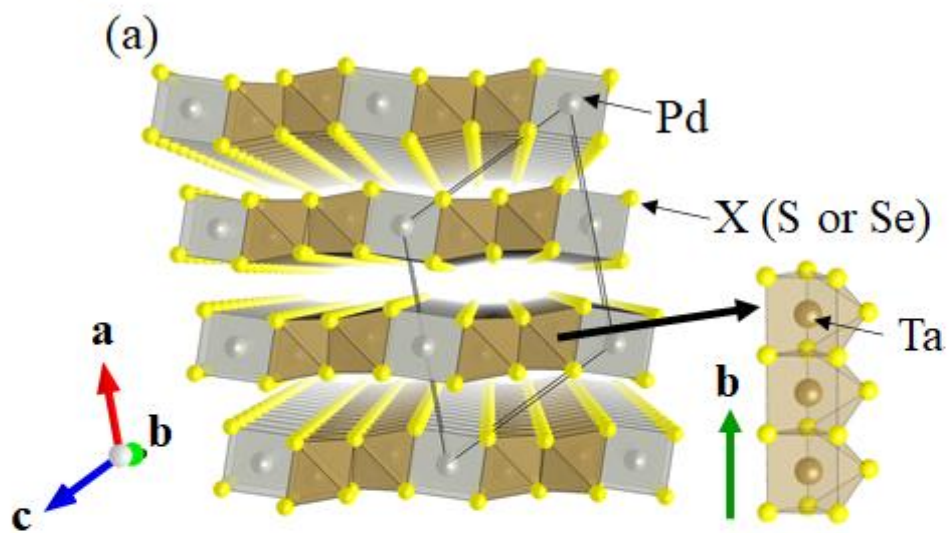

(b)Ta$_2$PdS$_6$

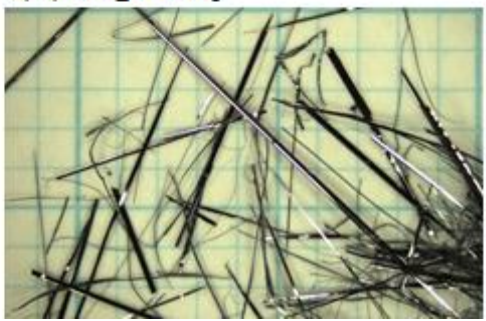

(c)Ta$_2$PdSe$_6$

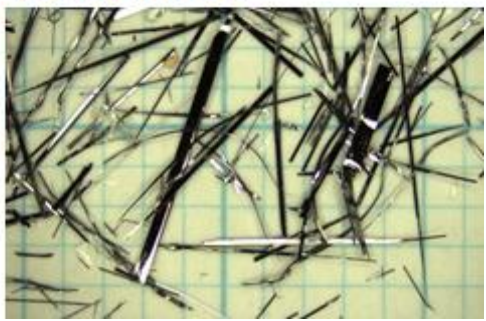

Figure 1. (Color online) (a) Crystal structure of Ta$_2$PdX$_6$. The right-side figure shows the TaX$_6$ prismatic coordination arranged along *b*-axis sharing the faces. Photographic images of single crystals of (b)Ta$_2$PdS$_6$ and (c) Ta$_2$PdSe$_6$.

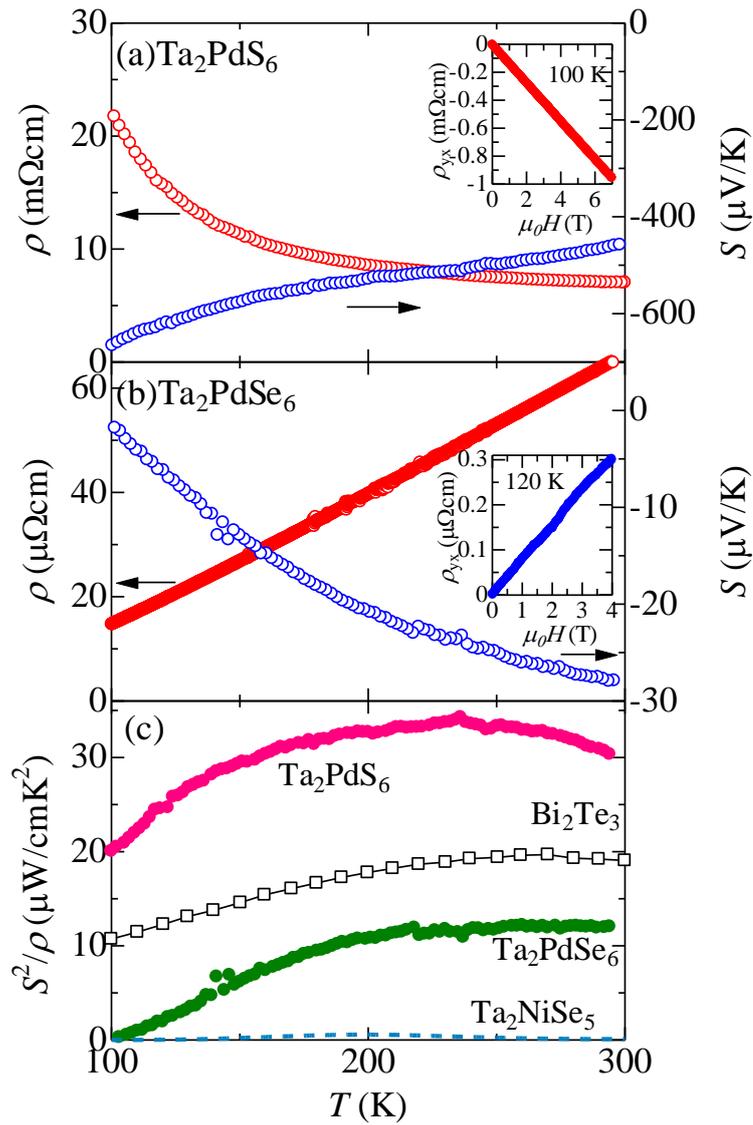

Figure 2. (Color online) Temperature dependence of the resistivity and the Seebeck coefficient of (a) $Ta_2PdS_6$ and (b) $Ta_2PdSe_6$. In the inset of (a) and (b) the Hall resistivity for $Ta_2PdS_6$ at 100 K and $Ta_2PdSe_6$ at 120 K are also shown. (c)Temperature dependence of the power factor $S^2/\rho$ of $Ta_2PdX_6$. The power factor of $Bi_2Te_3$[23] and $Ta_2NiSe_5$[15] is also shown for comparison.

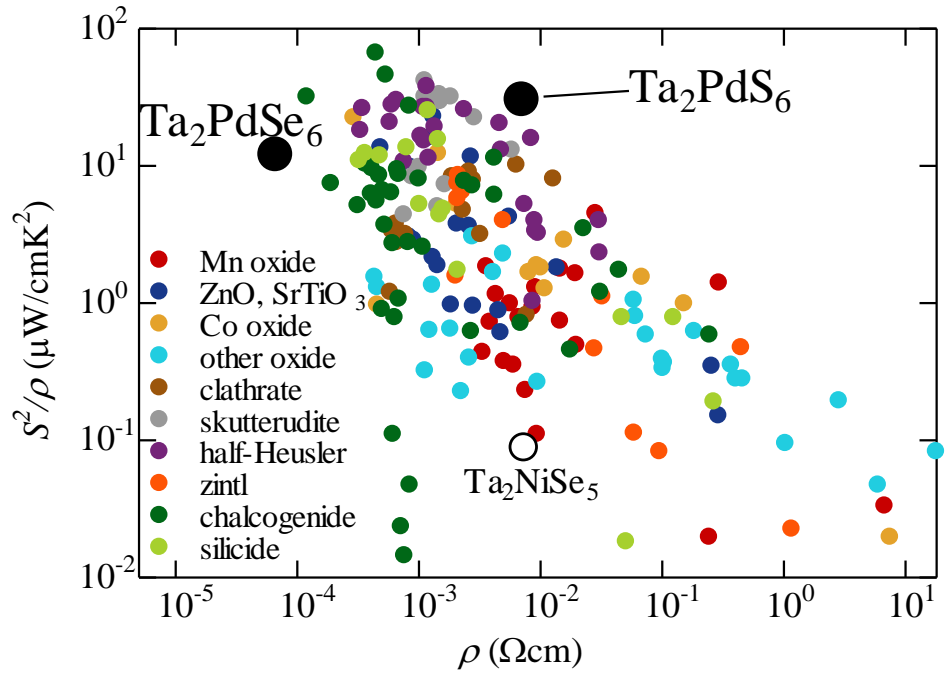

Figure 3. (Color online) Comparison of the power factor of $Ta_2PdX_6$(X=S, Se) with other thermoelectric materials. The data points except for $Ta_2PdX_6$ and $Ta_2NiSe_5$ are obtained from Materials Research Laboratory database of 300 K[24)].

Table 1. Summary of the used parameters to determine the thermal conductivity at 300 K.

| sample | Thermal diffusivity (m²/s) | Heat capacity (J/molK) | Density (mol/m³) | Thermal conductivity (W/mK) | ZT |
|---|---|---|---|---|---|
| $Ta_2PdS_6$ | $6.7\times10^{-6}$ | $2.2\times10^{2}$ | $8.5\times10^{3}$ | 14 | $6.4\times10^{-2}$ |
| $Ta_2PdSe_6$ | $8.9\times10^{-6}$ | $2.2\times10^{2}$ | $9.6\times10^{3}$ | 17 | $2.1\times10^{-2}$ |